\pdfoutput=1
\documentclass[aps,prd,nofootinbib,twocolumn,citeautoscript,showkeys,superscriptaddress]{revtex4-2} 

\usepackage{amsmath}   
\usepackage{graphicx}  % needed for figures
\usepackage{dcolumn}   % needed for some tables
\usepackage{bm}        % for math
\usepackage{amssymb}   % for math
\usepackage{epsfig}
\usepackage{epstopdf}  
\usepackage[utf8]{inputenc}
\usepackage{mdframed}
\usepackage{amsmath}
\usepackage{subcaption}
\usepackage{wasysym}
\usepackage[bookmarks=false]{hyperref} 
\hypersetup{pdfstartview=FitH,pdfhighlight=/O,colorlinks=true}
\usepackage[toc]{appendix}
\usepackage{color,soul} 
\usepackage{datetime}

\usepackage{enumitem}

\newcommand{\bea}{\begin{eqnarray}}
\newcommand{\eea}{\end{eqnarray}}
\newcommand{\ba}{\begin{eqnarray}}
\newcommand{\ea}{\end{eqnarray}}

\newcommand{\beq}{\begin{equation}}
\newcommand{\eeq}{\end{equation}}
\newcommand{\beqa}{\begin{eqnarray}}
\newcommand{\eeqa}{\end{eqnarray}}
\newcommand{\beqar}{\begin{eqnarray*}}
	\newcommand{\eeqar}{\end{eqnarray*}}

\renewcommand{\c}{$c$}

\begin{document}

 \title{Observational Opportunities for the Fuzzball Program} 
 \author{Daniel R. Mayerson}
 \email{daniel.mayerson@kuleuven.be}
  \author{Bert Vercnocke	} 
 \affiliation{Institute for Theoretical Physics, KU Leuven,
 Celestijnenlaan 200D, B-3001 Leuven, Belgium}
\date{\today}
%\keywords{}
%\pacs{}

\begin{abstract}
We discuss how string theory, and in particular the ``fuzzball'' paradigm, has already made and can make meaningful contributions to the phenomenology of strong gravity observations. We outline pertinent research directions for the near-future within this program, and emphasize the unique viewpoints that string theory and fuzzballs bring to phenomenology.
\end{abstract}

\maketitle

%%%%%%%%%%%%%%%%%%%%%%%%%%%%%%%%%%%%%%%%%%%%%%%%%%%%%%%%%%%%%%%%%%%%%%
%%%%%%%%%%%%%%%%%%%%%%%%%%%%%%%%%%%%%%%%%%%%%%%%%%%%%%%%%%%%%%%%%%%%%%
%%%%%%%%%%%%%%%%%%%%%%%%%%%%%%%%%%%%%%%%%%%%%%%%%%%%%%%%%%%%%%%%%%%%%%
%%%%%%%%%%%%%%%%%%%%%%%%%%%%%%%%%%%%%%%%%%%%%%%%%%%%%%%%%%%%%%%%%%%%%%
%\tableofcontents
\section{Introduction}

Recent years have seen a surge of developments in the study of black hole alternatives and potential ways in which one can observe deviations from the Kerr black hole paradigm of general relativity (GR). Such deviations may also contain hints to the underlying theory of quantum gravity.

In this work, we focus on black holes in string theory, and in particular as described by the fuzzball program \cite{Mathur:2005zp}. The main question that we wish to address is: what would a string theory black hole look like to an asymptotic observer --- or more precisely: what unique features would it display in strong gravity observations such as gravitational waves \cite{LIGOScientific:2016aoc} and black hole imaging \cite{EventHorizonTelescope:2022wkp}?
It has been argued many times before that string theory might offer insight into the structure of black hole microstates that statistically account for black hole entropy \cite{Strominger:1996sh,Mathur:2005zp,Bena:2013dka}. However, whether those microstates provide a way of seeing a testable deviation from the Kerr black hole is another matter. In particular, standard statistical physics arguments and holographic analysis suggest that the entropically favoured ``typical'' microstates would only differ at a Planck size from the horizon and would not directly be useful for experimental tests \cite{Balasubramanian:2005mg,Balasubramanian:2007qv,Balasubramanian:2008da,Raju:2018xue}. Nevertheless, we argue here that string theoretic black hole microstates give unique, novel, and important insights for strong gravity observations.

Fuzzballs are particular intricate states of branes and strings in string theory that have been argued to provide a microscopic description for black holes \cite{Mathur:2005zp}.
Motivated by the developments in understanding and constructing fuzzball solutions over the past years, we discuss here the most promising research directions for the near future; this includes identifying gravitational observables which will display non-trivial beyond-GR phenomena that are not excluded by the above statistical arguments.

The main outline of our arguments, suggestions, and motivations are threefold:
\begin{itemize}[leftmargin=*]
 \item From holography, we know that (semi-)classical solutions of string theory (that is: supergravity solutions) are extremely good in capturing long-range effects in strongly coupled quantum systems. Examples are confinement, gaugino condensation and the universality of the shear viscosity over entropy ratio $\eta = 1/4\pi$. Through holography, supergravity solutions offer precise connections with particle physics, hydrodynamics, condensed matter, and cosmology.
 
 If near-horizon microstate physics is similarly governed by strongly coupled degrees of freedom with long-range interactions dictating the fate of near-horizon low-energy excitations, then fuzzball solutions of supergravity, also known as ``microstate geometries'', offer great promise in giving a window into quantum gravity \cite{Bena:2007kg,Bena:2013dka,Bena:2022ldq}. The analogy is strengthened by properties of the solutions themselves, as microstate geometries share key features with the holographic gravity duals of strongly coupled systems, such as topological bubbles supporting fluxes and separation of scales described by long warped throats.
 
 In other words, microstate geometries provide a unique view on the necessary ingredients for horizon-scale microstructure in quantum gravity, and this leads to generic insights about the quantum nature of black holes and their near-horizon region. We explore this further in Section \ref{sec:general}.
 
 \item In the study of observables relevant for strong gravity experiments, we distinguish between \emph{macroscopic} and \emph{mesoscopic} observables.
 
The current catalogue of observed ultracompact astronomical objects is limited to black holes
and neutron stars. Do other stable objects exist that are more compact than a neutron star but
do not collapse into a black hole? If sufficiently compact and dark, these objects may have so
far evaded detection or have been mistaken as black holes \cite{Cardoso:2019rvt}.
 \emph{Macroscopic} observables are those that can distinguish such ultracompact objects from black holes (or neutron stars) in strong gravity observations. 
 The plethora of known microstate geometries are concrete metrics that can be made arbitrarily compact, and so offer a vast set of classical solutions which can be used as a probe of alternatives to horizons,\footnote{See also \cite{Guo:2017jmi} for related ideas.} regardless of their role in black hole quantum physics.

 Statistical arguments indicate that ``typical'' fuzzballs or horizon-scale microstructure may only differ at a Planck length from the naive black hole geometry \emph{in equilibrium}, and might be indistinguishable from a black hole horizon by any accessible macroscopic probe.
 However, quantum microstructure exhibits collective behaviour when \emph{perturbed significantly out of equilibrium} \cite{Brustein:2017nis,Kourkoulou:2017zaj,Brustein:2020tpg,Brustein:2021pof,Giddings:2017mym}. An analogy is the diffusion of particles in an otherwise featureless gas:
although we cannot see the individual gas particles directly, we can observe the consequences
of the underlying microscopic details in the Brownian jittering of the diffusing particles.
\emph{Mesoscopic} observables are then observables that are ``large'' enough to
measure in gravitational experiments, yet carry signatures from the collective behavior of the
underlying microscopics. They will point towards large \emph{quantum} effects during binary mergers which cannot be captured by classical geometries.

We explore macroscopic and mesoscopic observables further in Section \ref{sec:specific}, developing one particular example of each observable in some detail, and giving further general comments on mesoscopic observables.

\item There is a dearth of tools available that can be used for exploring and simulating various aspects of fuzzball dynamics. Most tools that simulate black hole imaging or gravitational wave signals are developed with general relativity (Kerr black holes) in mind, or at most theories that deviate minimally from GR. There is a need for the development of tools that can simulate aspects of fuzzball dynamics, if fuzzballs and string theory are to make meaningful predictions in dynamical processes. We discuss this in Section \ref{sec:tools}.

\end{itemize}

In this paper, we focus mainly on effects and observables that have not been directly discussed elsewhere. For an overview of earlier results and ideas in using fuzzballs in strong gravity phenomenology, we refer to \cite{Mayerson:2020tpn,Mayerson:2022yoc}; other related works include \cite{Bah:2021jno,Bacchini:2021fig,Bah:2020ogh,Bena:2022rna,Heidmann:2022ehn,Bah:2023ows}. The three following Sections \ref{sec:general}, \ref{sec:specific}, \ref{sec:tools} explore and discuss further our three main topics outlined above. We conclude in Section \ref{sec:conclusions}.

%%%%%%%%%%%%%%%%%%%%%%%%%%%%%%%%%%%%%%%%%%%%%%%%%%%%%%%%%%%%%%%%%%%%%%
%%%%%%%%%%%%%%%%%%%%%%%%%%%%%%%%%%%%%%%%%%%%%%%%%%%%%%%%%%%%%%%%%%%%%%
%%%%%%%%%%%%%%%%%%%%%%%%%%%%%%%%%%%%%%%%%%%%%%%%%%%%%%%%%%%%%%%%%%%%%%
%%%%%%%%%%%%%%%%%%%%%%%%%%%%%%%%%%%%%%%%%%%%%%%%%%%%%%%%%%%%%%%%%%%%%%
\section{General Predictions}\label{sec:general}

%%%%%%%%%%%%%%%
\subsection{Circumventing Buchdahl's Theorem} \label{sec:lovelockbuchdahl}
%%%%%%%%%%%%%%%

The Buchdahl bound \cite{Buchdahl:1959zz} is one of the most well-known results that is a key starting point in discussions of black hole physics beyond GR. In a sense, Buchdahl gives a measure of how ``difficult'' it is to construct objects that are of comparable compactness to black holes but do not collapse under their own weight to form a black hole.

Specifically, Buchdahl found that a spherically symmetric perfect fluid of total mass $M$ contained within a radius $R$, such that for $r>R$ there is only vacuum,  must have $R/M\geq 9/4 = 2.25$. Any radius smaller than this must lead to gravitational collapse to a black hole, which has $R/M=2$.
 There is a clear `gap' of allowed compactness for such horizonless solitons of matter, suggesting that drastic measures are necessary to circumvent this bound and resist the crushing pull of gravity --- such as violating energy conditions.

Constructing exotic compact objects (ECOs) in gravitational phenomenology can be seen as a study of how to circumvent Buchdahl's bound by relaxing its underlying assumptions \cite{Cardoso:2019rvt}.\footnote{Relaxing assumptions of the Buchdahl bound can also be studied in a more generic fashion, for example 	by considering mild anisotropy \cite{Andreasson:2007ck,Karageorgis:2007cy,Urbano:2018nrs} or non-perfect fluid elastic matter \cite{Alho:2022bki}. We focus here on specific known matter models that can violate the bound in some way.}
Perhaps the most well-known such objects are \emph{boson stars}, which are compact, self-gravitating solitons of a minimally coupled massive scalar field (possibly complex and possibly with higher-order potential terms). These boson stars can be made compact but not arbitrarily so: depending on the scalar field parameters, such as its mass, there is typically an upper bound on its compactness --- any scalar configurations that are more compact are unstable towards collapse into black holes.

A trade-off between compactness and stability is a common feature for many ECOs. On the other hand, microstate geometry solutions in string theory have shown us how solitons can be made that are arbitrarily compact --- in spirit, this means $R/M$ could be made arbitrarily close to the black hole value of $2$ for static vacuum black holes.

Microstate geometries must then evade Buchdahl's bound \cite{Mathur:2016ffb}. In particular, there is negative pressure along angular directions, and the energy density diverges near certain ``center'' singularities. However, these singularities are not pathologies of the solution, but simply an artifact of the restriction to four dimensions --- from the natural string-theoretic higher-dimensional point of view, these singularities are non-trivially resolved and the microstate geometry is completely smooth everywhere.

In fact, a remarkable result in the study of microstate geometries is that the \emph{only} way arbitrary compactness can be achieved without losing stability is by the existence of non-trivial topological structures (called ``bubbles'') in \emph{higher} dimensions that supports the matter from collapse \cite{Gibbons:2013tqa, Haas:2014spa,deLange:2015gca}; these topological ``bubbles'' are then kept stable from collapsing onto themselves by flux coming from particular string theoretic vector fields. Recently, it was argued that such microstructure can (always) be understood as bound states of fundamental ``themelia'' \cite{Bena:2022fzf}, extended objects in string theory that locally carry 16 supersymmetries.
The necessity of these topological structures and the additional string theoretic fields is an important feature of microstate  geometries and fuzzballs in general, and leads to generic predictions for the near-horizon physics that can potentially give rise to observable signatures. We discuss these further in the following Section.

\subsection{Predictions for scalars and vectors near horizon-scale microstructure}\label{sec:scalar_vector}

No-go theorems for solitons without horizons are evaded by microstate geometries, as discussed above. These solitons can be entirely smooth (free of singularities) as well as arbitrarily compact, but \emph{must} consist of non-trivial topology threaded by electromagnetic fluxes \cite{Gibbons:2013tqa,Haas:2014spa,deLange:2015gca}.

These necessary ingredients for microstructure have important implications on the necessary field content in the effective four-dimensional theory, which in turn can give rise to observable signatures of beyond-GR physics. In particular, the dimensional reduction of smooth microstate solutions to four dimensions necessarily include four-dimensional gauge fields and (pseudo-)scalars; these arise from the low-mass modes in the dimensional reduction of the metric and gauge fields in higher dimensions. 

These gauge fields and scalars must include a $\theta$-angle topological term of the form $\epsilon^{\mu\nu\rho\sigma}F_{\mu\nu} F_{\rho\sigma}$. Such terms are coupled to pseudoscalars or axions $\chi$, and potentially other scalar fields $\phi$. Schematically, the action contains terms such as:
\begin{equation}
f_{ij}(\chi,\phi) \epsilon^{\mu\nu\rho\sigma} F^i_{\mu\nu} F^{j}_{\rho\sigma}\,,
\end{equation}
where the $i$ runs over the number of vector fields, and $f_i(\chi, \phi)$ are (typically polynomial) functions of the (pseudo)scalars. 

The currently known black hole microstate models within string theory are all constructed in theories where these scalar fields are \emph{massless}, meaning they are moduli of the theory. Since no massless scalars have been observed, realistically these scalars must be given a (high) mass and the moduli must be \emph{stabilized}. As is well known in string cosmology, moduli stabilization is a tricky endeavour that usually implies extending the two-derivative Lagrangian with quantum or string effects \cite{VanRiet:2023pnx}.

The phenomenological importance of moduli stabilization seems to be in direct tension with the necessity of massless scalar fields supporting the non-trivial compact microstructure of black hole microstate geometries. However, in our view, the appearance of effectively (nearly-)massless scalars is precisely a \emph{prediction} of string theory black hole microstructure. In the highly compact (high-redshift) region, these axionic and other fields must be effectively massless to support the generic topological mechanisms underlying compact microstructure.
 This is irrespective of the mass at infinity that these fields obtain with moduli stabilization. In other words: the near-horizon region of generic fuzzball solutions should not strongly depend on the details of the moduli stabilization. This is reminiscent of the way in which constructions with a positive cosmological constant \`a la KKLT \cite{Kachru:2003aw} use the deformed conifold solution of \cite{Klebanov:2000hb} to describe local physics on the internal geometry.
 
The existence of fuzzballs then carries the prediction that near-horizon regions must be populated by nearly-massless scalar degrees of freedom. Very compact fuzzballs will be virtually indistinguishable from black holes surrounded by a tight cloud of scalar hair. This underscores the importance of studying hairy black holes as observationally relevant alternatives to Kerr; recent models of hairy black holes and their possible microstructure include \cite{Bah:2023ows}. Hair consisting of ultralight particles has already been shown to lead to possible observable consequences in gravitational wave observations, due to for instance superradiant instabilities and ``ionization'' resonances \cite{Baumann:2019ztm,Baumann:2022pkl}, dynamical friction \cite{Traykova:2021dua}, and altered finite-size properties (multipoles, tidal deformability) \cite{DeLuca:2022xlz}; see for example \cite{Bertone:2019irm} for an overview.

%%%%%%%%%%%%%%%%%%%%%%%%%%%%%%%%%%%%%%%%%%%%%%%%%%%%%%%%%%%%%%%%%%%%%%
%%%%%%%%%%%%%%%%%%%%%%%%%%%%%%%%%%%%%%%%%%%%%%%%%%%%%%%%%%%%%%%%%%%%%%
%%%%%%%%%%%%%%%%%%%%%%%%%%%%%%%%%%%%%%%%%%%%%%%%%%%%%%%%%%%%%%%%%%%%%%
%%%%%%%%%%%%%%%%%%%%%%%%%%%%%%%%%%%%%%%%%%%%%%%%%%%%%%%%%%%%%%%%%%%%%%
\section{Specific Predictions for Observables}\label{sec:specific}

In the previous section we focused on theoretical insights and generic predictions that arise from fuzzballs in string theory. These generic insights then lead to specific signatures and features in observations that can be searched for in gravitational wave and black hole imaging observations.
In this section, we explore examples of such signatures. 

As discussed in the Introduction, we distinguish between \emph{macroscopic} and \emph{mesoscopic} observables. The former type of observable can distinguish relatively diffuse compact objects that have macroscopic differences with black holes, whereas the latter type carries signatures of the microscopic, quantum nature of black holes.
Examples of macroscopic observables studied earlier in the context of fuzzballs (or other compact objects) include multipole moments \cite{Bena:2020see,Bianchi:2020bxa,Bena:2020uup,Bianchi:2020miz,Bah:2021jno,Fransen:2022jtw,Cano:2022wwo}, ringdown quasinormal modes \cite{Ikeda:2021uvc}, and shadows in black hole imaging \cite{Bacchini:2021fig,Cunha:2015yba,Cunha:2016bjh}.
As for mesoscopic observables, it was argued that tidal responses can encode quantum properties of black holes \cite{Brustein:2020tpg,Brustein:2021pof,Brustein:2021bnw,Chakraborty:2021gdf,Datta:2020rvo,Datta:2019epe} (see also Section \ref{sec:generalmeso} below); this also includes ``stringy'' tidal forces in fuzzballs \cite{Martinec:2020cml}. It is also possible that ringdown ``echoes'' may carry quantum signatures of the black hole microscopics \cite{Wang:2019rcf}.
(For a more comprehensive list of observables --- which can be either macroscopic or mesoscopic ---, and other references for the mentioned observables, see for instance \cite{Cardoso:2019rvt}.)

In the following subsections, we will give one example each of novel macroscopic and mesoscopic observable which has not been explored yet, and finally give some general comments on mesoscopic observables.

%%%%%%%%%%%%%%%
\subsection{Macroscopic example: Imaging photon rings}\label{sec:imaging}
%%%%%%%%%%%%%%%

Very-Long-Baseline Interferometry (VLBI) such as the Event Horizon Telescope \cite{EventHorizonTelescope:2022wkp} and future successors \cite{Blackburn:2019bly,Haworth:2019urs} are able to ``image'' black holes by capturing light from its environment of a black hole. Before reaching us, this light first travelled close to the black hole and experienced its strong gravitational pull, and so encodes features of the black hole geometry.

\emph{What if the object being imaged is not a black hole but rather a compact, horizonless object?} Would we be able to tell the difference in VLBI observations?

\begin{figure*}[htbp]
\begin{subfigure}{0.48\textwidth}
		\includegraphics[width=\textwidth]{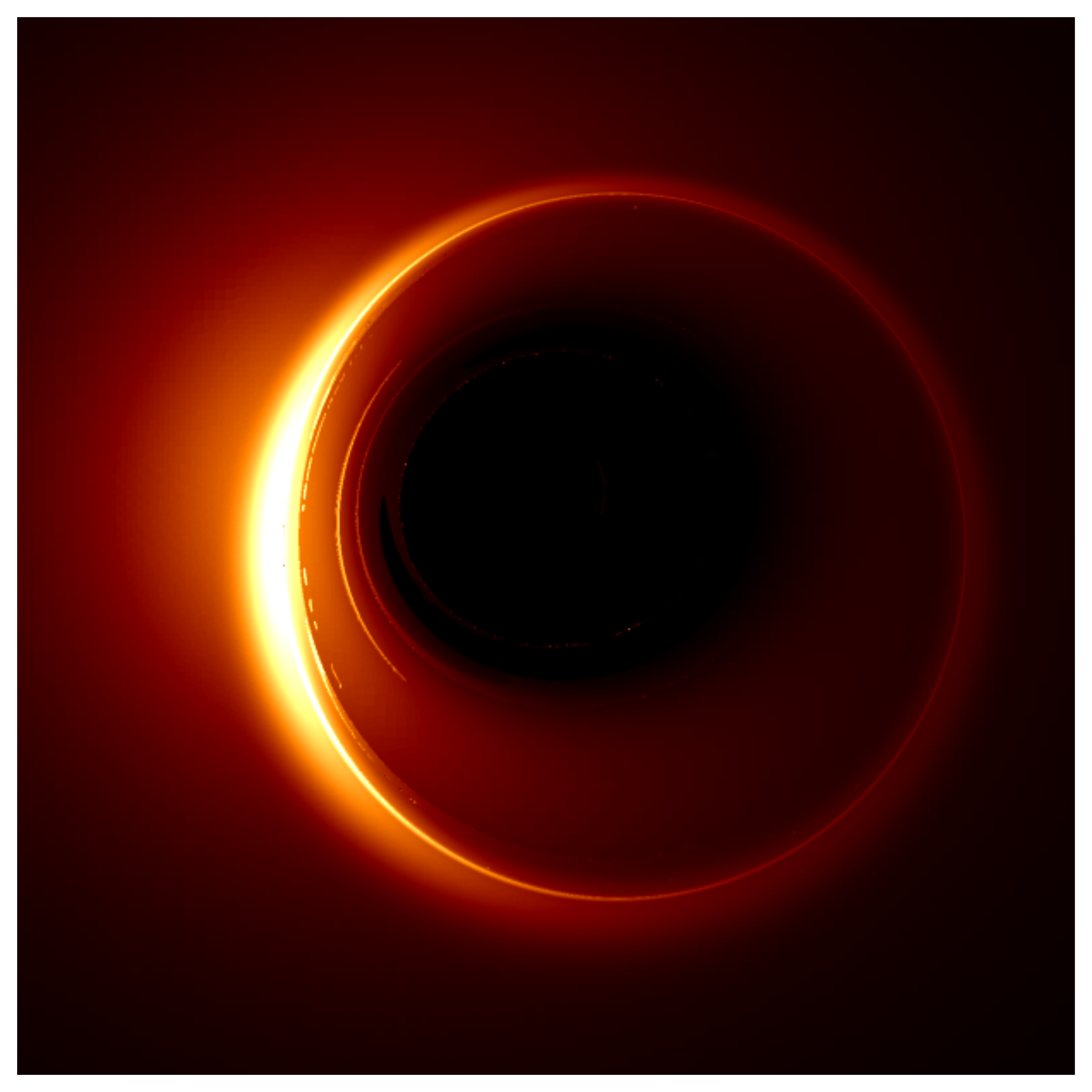}
		\caption{Manko-Novikov black hole image	}
\end{subfigure}
\begin{subfigure}{0.48\textwidth}
		\includegraphics[width=\textwidth]{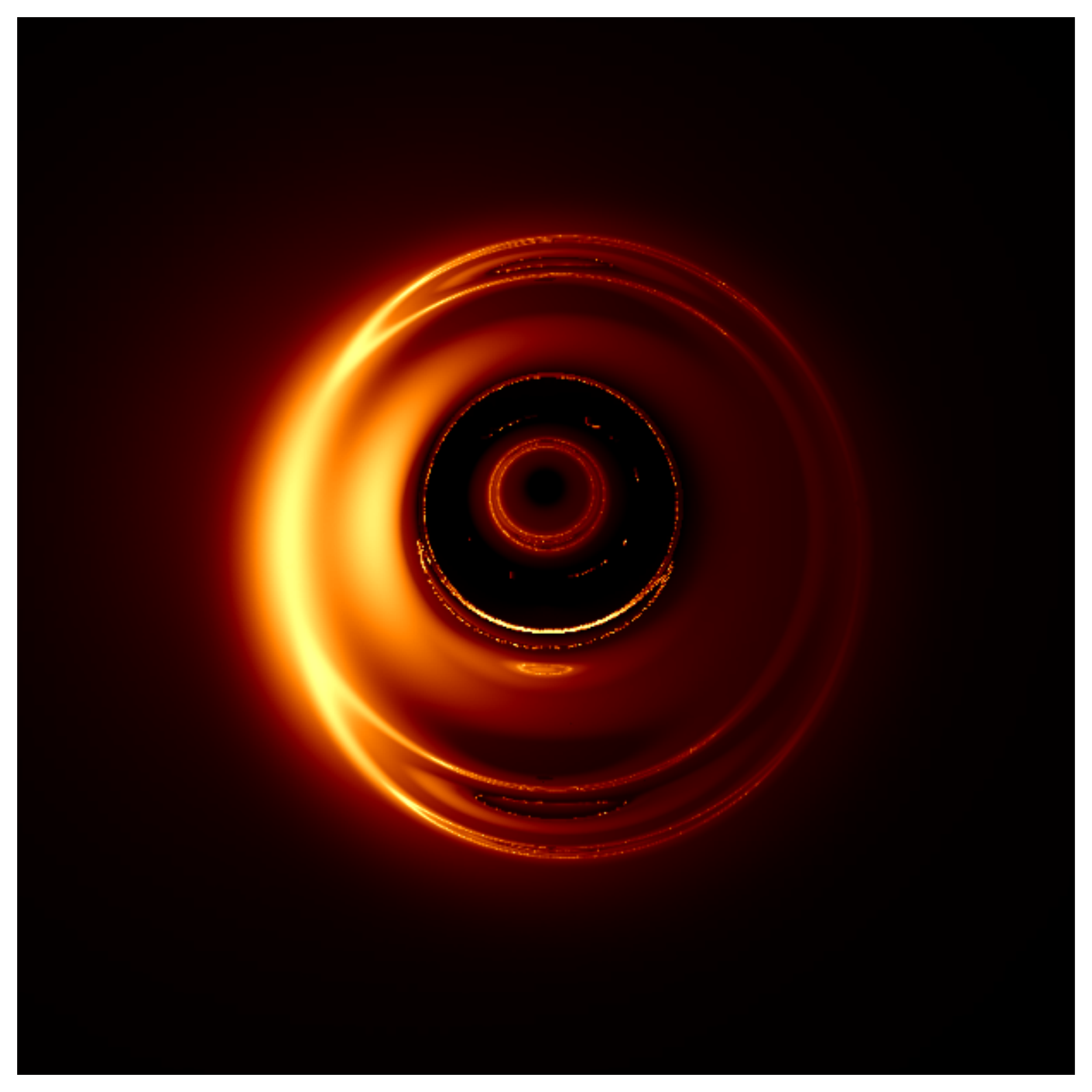}
		\caption{Fuzzball image}
\end{subfigure}

\caption{\raggedright Simulated images of a Manko-Novikov black hole and a fuzzball, both imaged at an inclination $\theta_0=30^\circ$. The bright photon ring surrounds the central dark shadow region. In the Manko-Novikov black hole, a sharp secondary ``ring'' structure is seen on the inside of the primary photon ring (on the left side of the image); this is a typical signature of chaotic photon trajectories. The fuzzball displays multiple such secondary photon ring structures in its image.
\newline
Figures made with \texttt{FOORT} \cite{FOORT}. The Manko-Novikov black hole has parameters $J/M^2=0.94$ and $\alpha_3=2$ (see \cite{Staelens:2023jgr} for details). The fuzzball is the ``ring fuzzball'' introduced in \cite{Bacchini:2021fig} with parameters $P=2,q_0=50$ and $\lambda = 0.19$. The emission is simulated according to the model described in \cite{Cardenas-Avendano:2022csp} (see also \cite{FOORT}) with parameters $\xi=\beta_r=\beta_\phi=1, \gamma=-1.5$ and $\mu=0.66,\sigma=0.5$ (for Manko-Novikov) or $\mu=-0.5,\sigma=2.0$ (for the fuzzball).}
\label{fig:BHandFB}
\end{figure*}

The most obvious effect one can imagine is the absence of the shadow or dark depression central to the black hole image (see Fig. \ref{fig:BHandFB}) --- if there is no longer a horizon to absorb that can absorb light rays, the object is in principle completely \emph{transparent}. This was certainly seen in studies of boson star images \cite{Cunha:2015yba,Cunha:2016bjh}.
However, as we showed in \cite{Bacchini:2021fig}, fuzzballs --- even though they are horizonless --- \emph{do} mimic the degree of darkness of a black hole shadow when the effects of redshift, timing and tidal forces (curvature) are correctly taken into account on the photons that explore the horizon-scale microstructure.
In the end, these effects give fuzzball images an effective shadow and an indistinguishable image as compared to a black hole \cite{Bacchini:2021fig}.\footnote{Of course, another reason for a dark depression in the image can be due to lower emitted flux in the central region; see for example \cite{Vincent:2015xta,Olivares:2018abq} for realistic boson star accretion models.}

The main take-away from the emergent fuzzball shadows of \cite{Bacchini:2021fig} is that fuzzballs can show us important and non-trivial features of imaging self-consistent and arbitrarily compact horizonless objects. Fuzzballs are quite unique in this regard. For example, imaging studies of boson stars \cite{Cunha:2015yba,Cunha:2016bjh} revealed similar ``internal'' chaotic structures in the images, but could not see the emergence of shadows as these objects seem to lack the necessary properties (sufficient compactness, and especially the resulting redshift and tidal forces that the geodesics experience).

\medskip

\begin{figure}[htbp]\centering
		\includegraphics[width=.45\textwidth]{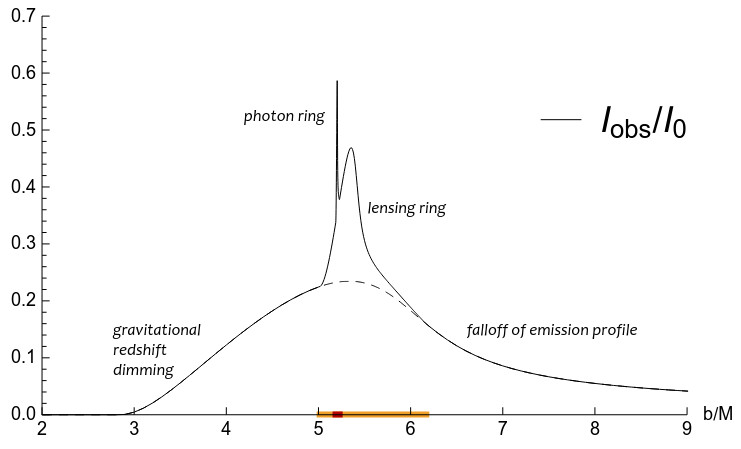}
		\caption{Normalized luminosity vs.\ impact parameter $b$, in a Schwarzschild background with mass $M$. Picture from \cite{Gralla:2019xty}}
\label{fig:brightness}
\end{figure}

Properties of the image sourced by the surrounding plasma, more than the dark shadow region contained within, can reveal beyond-GR features.
For Kerr black holes, one can characterize geodesics by the number of times $n$ they cross the equatorial plane before reaching the observer \cite{Gralla:2019drh}. The photons that travel on orbits crossing the equatorial plane a number of times give rise to a series of concentric rings in the image, collectively called the photon ring (see also Fig. \ref{fig:BHandFB}).
% \cite{Luminet:1979nyg,Hollywood:1997jkq,Beckwith:2004ae}. 
The luminosity profile for emission from an optically thin disk has a distinctive multi-peak structure \cite{Gralla:2019xty}: the primary peak is sourced by the $n=1$ light-rays (the ``lensing band''), while for $n\geq 1$ one observes exponentially smaller and higher peaks; see Fig. \ref{fig:brightness}. 
The higher-order ($n>0$) rings encode features of the underlying geometry and have been argued not to depend on the details of the astrophysical emission source; these higher rings may be observable with future space-based VLBI \cite{Gralla:2020srx,Paugnat:2022qzy}. Examples of photon ring features that encode properties of the geometry include the \emph{shape} of the photon ring \cite{Gralla:2020srx} and critical exponents for null geodesics --- such as the Lyapunov exponent, which characterizes the rate at which successive photon subrings decrease in width \cite{Johnson_2020, hadar2021photon}. These features have been argued to be potentially measurable with near-future, plausible VLBI observations \cite{Johnson_2020, hadar2021photon, Gralla:2020srx}.

Most studies of observations have focused on calculating photon ring properties for Kerr and understanding the possible precision to which we can expect to measure them in (near-)future observations.
However, the question of which of these properties can be used to distinguish Kerr from alternatives --- be it other, non-GR black holes or horizonless compact objects --- remains largely unexplored. A first study into how well the Lyapunov exponent and photon ring shape can distinguish between Kerr and alternative black holes that deviate from Kerr was performed in \cite{Staelens:2023jgr}, although many important questions remain.

For example, it is not clear what the global structure of the photon rings is expected to be in general, ultracompact object spacetimes. When null geodesics are no longer separable, the null geodesics exhibit chaotic behaviour and one general feature appears to be that the critical curve (sometimes called ``shadow boundary'') will no longer be a connected curve \cite{Staelens:2023jgr, Cunha:2016bjh}. Since the higher order photon rings are supposed to approach the critical curve more and more, this also suggests that the photon rings should also show disconnected features for such spacetimes. This feature becomes apparent when imaging accretion around fuzzballs, but also beyond-GR black holes where null geodesics are not separable. We illustrate this for both a fuzzball and a separability-breaking black hole in Fig. \ref{fig:BHandFB}.

In general, it is an open problem to investigate in what ways the intensity peak structure as in Fig. \ref{fig:brightness} for a compact object image will be similar or different to that of a (Kerr) black hole.\footnote{See \cite{Cunha:2017qtt,Cunha:2020azh} for results on light ring existence and properties for generic (compact) objects.}
Fuzzballs have parameters that can be dialed to increase compactness at will (the so-called ``scaling'' limit); this makes them the ideal tool to ask questions such as: How compact does an object need to be before photon (sub)rings exist in its images? When and if the rings exist, does their relative thickness (set by the Lyapunov exponent) and other properties depend strongly on the compactness? In general, how compact can an object be before it becomes indistinguishable from a black hole in imaging observations?
These questions motivate an overarching program to map out robust plasma-independent observables (such as the photon ring shape and critical exponents) that characterize images of compact objects and distinguish them from Kerr images.

This program clearly requires tools that can simulate (realistic) photon emission from accretion discs. Until recently, most image-simulating ray tracers focus on the Kerr metric and are difficult to customize to other metrics. The recently developed ray tracer \texttt{FOORT} \cite{FOORT} was developed with the explicit goal to make implementing arbitrary metrics as easy as possible, \emph{and} includes realistic emission models that facilitate mock observational analyses of beyond-GR objects (see Fig. \ref{fig:BHandFB}). Developing such tools are a crucial step towards the high-energy community making contact with the observational world; we discuss this thread further in Section \ref{sec:numerical}.

\subsection{Mesoscopic example: Gravitational wave transitions in binary mergers}\label{sec:quantumdynamics}

Can one detect actual \emph{quantum} effects of stringy black holes? The underlying quantum state of a probed black hole is expected to behave (quasi-)classically by decoherence and/or eigenstate thermalization. Fine-grained, microscopic observables that can resolve the precise quantum state are inaccessible to the low-energy gravitational wave (or black hole imaging) observations, while coarse grained observables of macroscopic properties such as the black hole mass or spin will not be able to distinguish between different quantum states.

\emph{Mesoscopic} observables --- which should show a signature of the quantum nature of the state but still be accessible in observations --- are signals coming not from the quantum state itself, but from \emph{transitions} between states. One would expect such signals to be the largest during the merger phase of a binary system. 
At this point, the merging black holes are in a highly excited state, which could lead to intermediate transient states in the evolution, in analogy to resonances in particle physics. Such intermediate states correspond to atypical quantum states, such as highly non-generic classical fuzzball geometries. 
The transitions associated with such intermediate quantum resonances are expected to give different effects from the numerical relativity prediction of the merger evolution.

Quantum tunneling processes between intermediate states may lead to a different overall speed than the numerical (classical) relativity prediction for the merger. 

For example, the final merger may happen faster than expected if certain low-probability tunneling ``shortcut'' paths are taken in the evolution.  Indeed, the transition time between particular microstates is expected to be very short $\tau \sim e^{-S/\hbar}$ (with $S$ the Bekenstein-Hawking entropy of the black hole  \cite{Mathur:2008kg,Kraus:2015zda,Bena:2015dpt}).
On the other hand, if the tunneling procedure ends up in a meta-stable ``glassy'' state near the classical (stable) final state \cite{Anninos:2011vn,Charles:2018oob}, this could be longer-lived and then there would be less radiated energy than expected in the merger process.

Transitions between intermediate resonances can also be paired with energy exchange with the environment, which can happen through several channels. First, the transition can be associated with a burst of gravitational waves, for which one can obtain an order of magnitude estimate from the change in multipoles going from a highly non-generic fuzzball to a typical Kerr-like end state \cite{Hertog:2017vod}. Second, more intriguingly, also the axion-vector support that upholds the solution as discussed in section \ref{sec:lovelockbuchdahl} will be involved in the transition. If the supporting axion-vector matter is not merely stuck in a hidden sector but has couplings to standard model matter, the energy exchange of the transitions can also give rise to a potentially detectable non-gravitational counterparts of the merger in the electromagnetic and/or neutrino range.
Such counterparts would be of a different nature from the (known) axion or other boson clouds around black holes or fuzzballs; as discussed in Section \ref{sec:scalar_vector}, such clouds are usually highly localized ``hair'' with minimal long-range effects.

An example of how one could detect such signals is to look for ``missing'' energy during the merger process compared to the GR prediction --- such energy must then have been imparted into the additional matter sectors. For example, for binary component masses $M_1,M_2$ (as measured from the inspiral phase) and a final mass $M_{\rm end}$ of the merger black hole (as measured from the ringdown phase), we obtain the radiated mass $(\Delta M)^{\rm (obs)} = (M_1+M_2)-M_{\rm end}$. This can be compared to the prediction from numerical relativity\footnote{Packages to extract remnant black hole properties from a merger have been developed, see \cite{Islam:2023mob} for an example.}  (given the same initial masses $M_1,M_2$), $(\Delta M)^{\rm (GR)}(M_1,M_2)$; one could then analyze whether there is a statistical significant difference between $(\Delta M)^{\rm (obs)}$ and $(\Delta M)^{\rm (GR)}$. Such analyses will especially become interesting when stacking a large number of very precise measurements, such as will be available with the third generation of gravitational wave detectors.

More generically, quantum effects may lead to a violent merger phase which differs from the (numerical) relativity expectation. 
This motivates searches to match the waveforms of the inspiral and ringdown phases separately. Any mismatch of such separated modelling compared to the full numerical relativity waveform would hint at beyond-GR effects.

%For component masses $M_1,M_2$ obtained from the inspiral phase, and a mass  $M_{\rm end}$ of the black hole after merger, obtained from ringdown analysis, one should leave open the possibility that we have a detectable mass difference $\Delta M = M_{\rm end}- (M_1 + M_2)$ compared to the best numerical data fit over the entire waveform. At the time of writing, the observational bounds are not stringent enough to hint at such a difference. It would be very interesting to see whether this can be put to the test in the future, especially with the precision and many expected observations of third generation gravitational wave detectors. By binning the results according to the inspiral data (component masses and spins, details of infalling orbit), one can write down an average $M_{\rm end}$ obtained from for each bin. Then one can consider the variance by averaging over the bins: $\sigma_M = \langle M_{\rm end}^2\rangle-\langle M_{\rm end}\rangle^2$. The bins should be chose such that $\sigma _M =0$ for GR merger processes. Quantum transitions involving fuzzballs then predict a non-zero $\sigma_M$.

\subsection{Comments on mesoscopic observables}\label{sec:generalmeso}

The example discussed above illustrates the key feature of any mesoscopic observable: it is a consequence of exploring the \emph{phase space} of black hole microstates, and gives deviations from classical GR precisely because this phase space has non-trivial, classically unexpected properties.
It is important to emphasize that this phase space exploration is a \emph{necessity} for mesoscopic observables, since a \emph{single} typical microstate should not exhibit noticeable deviations from the thermal average black hole geometry \cite{Balasubramanian:2005mg,Balasubramanian:2007qv,Balasubramanian:2008da,Raju:2018xue}.

The non-trivial quantum phase space is also the heart of the argument how collapsing matter can avoid formation of horizons, even without large curvature. When the collapsing matter becomes approximately horizon-sized, a \emph{large phase space} of possible states (fuzzballs) becomes available for the matter to tunnel into. The probability of tunneling into any given state is exponentially small: $\mathcal{O}(e^{-S})$, where $S$ is the entropy of the matter; but precisely because there are $e^S$ microstates available, the probability of tunneling into \emph{any} (horizonless) fuzzball state becomes $\mathcal{O}(1)$ \cite{Mathur:2008kg,Kraus:2015zda,Bena:2015dpt}.

Any mesoscopic observable must then be related to a non-trivial property of the quantum phase space of fuzzballs. Of course, this makes finding and quantifying such observables a hard problem, as by definition they lie outside any semi-classical supergravity description and are not described by microstate geometries.

Another exciting mesoscopic observable that has been suggested is \emph{tidal Love numbers}; these quantify the response of an object to an external gravitational tidal force. For a single geometry that represents a typical black hole microstate, these tidal Love numbers will necessarily be very close to the corresponding black hole values \cite{Balasubramanian:2007qv, Bianchi:2022qph}.\footnote{Holographically, the tidal Love number is related to four-point functions such as $\langle H LL H\rangle$, where $H$ (resp. $L$) is a heavy (resp. light) operator; such four-point functions can be calculated as two-point functions of a light probe ($L$) on a geometry that corresponds to $H$. In such a context, it was shown that a typical microstate geometry will indeed exhibit two-point functions that are indistinguishable from the black hole two-point functions until very late times \cite{Balasubramanian:2007qv, Bena:2019azk}. However, exploring the phase space precisely means one should also take into account all of the possible correlators of the form $\langle H LL H' \rangle$ with $H\neq H'$. Such ``off-diagonal'' correlators are a largely unexplored topic in holography due to their technical difficulty (see also \cite{Dimitrov:2021csq}).}
However, it was argued in \cite{Brustein:2021bnw,Brustein:2020tpg,Brustein:2021pof} that tidal forces can induce long-range \emph{collective resonances} over the phase space of microstates, giving rise to effective non-trivial tidal Love numbers that deviate from their corresponding black hole expectation and could in principle be measured. Again, the contribution from the entire phase space of microstates is crucial in this argument.

%%%%%%%%%%%%%%%%%%%%%%%%%%%%%%%%%%%%%%%%%%%%%%%%%%%%%%%%%%%%%%%%%%%%%%
%%%%%%%%%%%%%%%%%%%%%%%%%%%%%%%%%%%%%%%%%%%%%%%%%%%%%%%%%%%%%%%%%%%%%%
%%%%%%%%%%%%%%%%%%%%%%%%%%%%%%%%%%%%%%%%%%%%%%%%%%%%%%%%%%%%%%%%%%%%%%
%%%%%%%%%%%%%%%%%%%%%%%%%%%%%%%%%%%%%%%%%%%%%%%%%%%%%%%%%%%%%%%%%%%%%%
\section{The Necessity of Developing Numerical Tools}\label{sec:tools}
%\subsection{Developing tools: The necessity of numerical methods}
\label{sec:numerical}
%%%%%%%%%%%%%%%

Many observable consequences of beyond-GR features in black hole physics can only be revealed by understanding the \emph{dynamics} of black hole-replacing objects. This remains one of the main obstacles that lie between string theory with its insights into compact black hole microstructure, and being able to make specific, quantitative predictions for new phenomena in observations.

There are many challenges that need to be addressed to overcome this obstacle. In particular:
\begin{itemize}
 \item[(i)] Most of the microstate geometries constructed are supersymmetric or extremal, implying they need to be somehow perturbed before any dynamics can be studied.
 \item[(ii)] As also discussed in Section \ref{sec:lovelockbuchdahl}, these solutions necessarily include a number of additional gauge fields and scalars that couple non-trivially to the metric field; this adds a layer of complexity to dynamical questions.
 \item[(iii)] It is not even clear how well the effective classical (super)gravity action captures the correct physics of microstate geometries; arguments first put forward by Kraus \& Mathur suggest that quantum tunneling effects can become important in the evolution of a collapsing system (and invalidate the classical evolution picture) well before the curvature scales become large \cite{Kraus:2015zda,Mathur:2008kg,Mathur:2009zs,Mathur:2016ffb,Bena:2015dpt}.
\end{itemize}

Addressing challenge (i), where the supersymmetric nature of many microstate geometries implies a lack of interesting dynamics, can be addressed in various ways. The most pertinent resolution would be to construct explicit non-supersymmetric and non-extremal smooth microstate geometries, which are more suitable for dynamical questions. Examples of such constructions can already be found in approaches based on perturbing known supersymmetric solutions (superstrata) into non-supersymmetric ones (dubbed ``microstrata'') \cite{Ganchev:2021pgs}, or in the approach of \cite{Bah:2020ogh,Bah:2021owp,Bah:2022yji}, where clever mathematical structures within the supergravity equations of motion are exploited.

\medskip

To address challenge (ii), the approach that has been most explored so far is to focus on dynamical questions where backreaction can be ignored. This is warranted if one only focuses on a particular sector of the full dynamics, which is then assumed to decouple from the rest. For example, a minimally coupled scalar field perturbation on top of a microstate geometry background was used to calculate the scalar ringdown dynamics (quasinormal modes and late-ringdown ``echoes'') of particular microstate geometries \cite{Ikeda:2021uvc}.

Another example within this approach is simulating black hole imaging phenomena. This is done by ray-tracing null geodesics on geometries, assuming that the imaging photons do not couple to the non-gravitational fields of the solutions. However, most ray tracers are geared mainly towards implementing Kerr as the background black hole metric, and there is often a high barrier to implementing other metrics.
\texttt{FOORT} \cite{FOORT} is a recent ray-tracer that was developed specifically to make it as easy as possible to ask observationally relevant questions for arbitrary metrics --- this includes high-resolution imaging and the production of simulated interferometric visibility amplitudes, which is what is measured in Very-Long-Baseline Interferometry such as the Event Horizon Telescope. \texttt{FOORT} has already been used to understand whether black holes that deviate from Kerr can show observable differences in their photon ring shape and Lyapunov exponent \cite{Staelens:2023jgr} (see also Section \ref{sec:imaging}). \texttt{FOORT} is thus a recent example of a tool development that is aimed explicitly at connecting string theory beyond-GR black hole models with observations in the context of black hole imaging.

Realistic tools that can simulate gravitational wave signals from binary mergers are necessarily more involved then the above approaches, where only the effect of the non-dynamical curved spacetime is considered. Nevertheless, besides attempting a complete numerical solution of complicated microstate geometry dynamics, there are other possible ways forward in attacking such questions.
One possible technique is that of reduction of order (as discussed in for instance \cite{Allwright:2018rut}). Here, one arranges the complete differential equation into a leading part and a subleading source $S$:
\begin{equation} 
	\label{eq:GSeq} G(g_{\mu\nu}) = \epsilon\, S(g_{\mu\nu}), 
\end{equation}
where $G$ is the Einstein tensor and $S$ represents the energy-momentum tensor.\footnote{This is a natural split of the Einstein equations when the matter fields are expected to be ``small'' compared to the metric curvature scales.}
 One then solves this differential equation iteratively, order by order in $\epsilon$: $G(g_{\mu\nu}^{(0)}) = 0$, $G(g_{\mu\nu}^{(1)}) = \epsilon\, S(g_{\mu\nu}^{(0)})$, etc. As long as $|g_{\mu\nu}^{(i+1)}| < \epsilon\, g_{\mu\nu}^{(i)}$, this perturbation scheme remains valid. Such an approach is well-suited to solve for higher-derivative corrections to GR, which exhibit a natural split as in (\ref{eq:GSeq}) into a subleading source term containing the higher-derivative corrections \cite{Endlich:2017tqa}.\footnote{See also \cite{Cayuso:2023aht} for a alternative recent approach that is promising to numerically integrate higher-derivative corrections.}

Such an approach may give us answers to natural questions in fuzzball dynamics, such as whether it is possible to reach one of the many smooth, horizonless microstate geometries by (semi)classical evolution from collapse or merger scenarios. Extending gravitational dynamics beyond GR to theories with scalars has been recently attempted in various works \cite{Benkel:2016kcq,Benkel:2016rlz,Okounkova:2017yby,Witek:2018dmd}; from these, gravitational collapse without matter or with only certain scalar fields has been shown numerically to lead to horizon formation \cite{Malafarina:2017csn}, including possibly relaxation to black holes with scalar hair \cite{Benkel:2016rlz}, or alternatively to oscillating boson star configurations \cite{Seidel:1993zk}. 
Techniques such as iterative reduction of order may give insight into whether more complicated theories of matter --- such as those that include smooth microstate geometries --- can also have alternative evolution scenarios into such horizonless geometries.

\medskip

Finally, addressing (iii) is the most difficult challenge, as by definition it takes us outside of the validity regime of (classical) supergravity theories.
There can still be hints from classical dynamics that indicate where such quantum effects can become important. In the reduction of order approach mentioned above, precisely where the perturbative solution breaks down could hint at the emergence of new, stringy physics. For example, microstate geometries consist of a number of ``centers'', which in the higher dimensional picture are points at which a compact circle pinches off to zero size --- in the effective four-dimensional picture, these centers are singularities. The breaking down of the evolution of a perturbative solution in a (super)gravity theory that contains such multicentered geometries can indicate the creation or formation of such a center.

However, as we emphasized above in Section \ref{sec:generalmeso}, it is precisely the effects that explore the non-trivial quantum phase space of gravity that lead to the most interesting (mesoscopic) effects --- such as quantum tunneling avoiding the formation of horizons \cite{Kraus:2015zda,Mathur:2008kg,Mathur:2009zs,Mathur:2016ffb,Bena:2015dpt}. By definition, these effects do \emph{not} show up in a conventional breakdown (such as diverging curvature) of the classical theory, and it is not clear how classical dynamics could give any indications for their presence. Incorporating them into dynamical simulations, even in principle, remains a wide-open question.

%%%%%%%%%%%%%%%
\section{Conclusions}\label{sec:conclusions}
%%%%%%%%%%%%%%%

We have hinted at possible avenues to learn from the intersection of the fuzzball program and current efforts to observe black holes.
Even though no arbitrarily compact microstate geometries exist with the same asymptotic charges as the Kerr or Schwarzschild black holes, by studying the structure of the solutions currently available, be they charged or higher-dimensional, one learns new ways to understand four-dimensional black holes of GR as well.

Fuzzballs are not the only approach within string theory to resolve the information paradox and describe the quantum nature of black holes. Other approaches include for instance non-violent non-locality \cite{Giddings:2006sj,Giddings:2012bm,Giddings:2012gc} or islands \cite{Almheiri:2019psf,Almheiri:2019hni,Penington:2019npb}, where the non-locality of quantum gravity is central. It would certainly be interesting to understand how fuzzballs are related to these approaches, what the different approaches can learn from each other, and the possible observational consequences that can be derived from such alternative approaches.

To summarize, the three main take-aways from our discussion are:
\begin{itemize}[leftmargin=*]
 \item String theory explicitly motivates studying hairy black holes as effective models of black hole microstructure, since generically new effectively massless scalar fields are expected to emerge in near-horizon physics (Section \ref{sec:scalar_vector}).
 \item Even if a ``typical'' quantum state of a black hole is indistinguishable from the classical black hole geometry in equilibrium, \emph{non-equilibrium} dynamics  encoded in \emph{mesoscopic} observables that reflect features of the non-trivial \emph{quantum phase space} can reveal underlying quantum microstructure (Sections \ref{sec:quantumdynamics} \& \ref{sec:generalmeso}, and the Introduction). 
 
 \item To make explicit contact with observational data, we need access to fuzzball dynamics --- specifically, the classical evolution of microstate geometries in supergravity, but also the quantum dynamics on the non-trivial phase space. This requires developing further simulation tools (Section \ref{sec:tools}).
 
\end{itemize}

Specifically, we have identified a number of exciting and pertinent research questions and directions for fuzzball research to make further contact with observations:
\begin{itemize}[leftmargin=*]
 \item  Construct an explicit, smooth horizonless microstate geometry which has its moduli stabilized at infinity but where the scalars become effectively massless in the high-redshift region near the microstructure (Section \ref{sec:scalar_vector}).
 \item  On the one hand, highly compact fuzzballs behave much like hairy black holes, giving a strong theoretical motivation to study hairy black holes in an observational context.
 On the other hand, dark matter fields that are light or massless give rise to a wide variety of possible observable effects. 
  Which dark matter effects are applicable to the specific matter fields predicted by black hole microstructure, where the matter is effectively massless \emph{only} at the horizon scale, but massive elsewhere (Section \ref{sec:scalar_vector})?  For example, if the microstructure hair interacts with standard model fields, it can give rise to unexpected electromagnetic counterparts to black hole mergers (Section \ref{sec:quantumdynamics}).

 %%%%
 \item  How does the absence of a horizon in (ultra)compact object affect the photon (sub)ring properties? Are effects of the internal, compact structure, such as the discussed chaotic photon trajectories, observable in the interferometric signatures of the photon rings (Section \ref{sec:imaging})?
 %%%%
  \item  How can we obtain additional insight into the possible stringy or quantum effects (and their magnitude) that can be observed in the highly non-linear merger phase of binary mergers? For example, is there a statistically significant difference between the merger energy loss as observed, $(\Delta M)^{\rm (obs)}$, and predicted by general relativity $(\Delta M)^{\rm (GR)}$ (Section \ref{sec:quantumdynamics})?
 %%%%
  \item  Develop a numerical scheme (using reduction of order or otherwise) that can simulate dynamical evolution and formation of horizonless microstate geometries in complicated supergravity theories. Can breakdowns of these numerical methods be used to understand formation of stringy phenomena such as higher-dimensional ``centers''? Can one reach a classical fuzzball end state in a numerical scheme of dynamics (Section \ref{sec:tools})?
\end{itemize}

%%%%%%%%%%%%%%%%
\acknowledgments

We thank Fabio Bacchini, Iosif Bena, Nikolay Bobev, Jan de Boer, Ram Brustein, Vitor Cardoso, Lies Van Dael, Thomas Hertog, Lorenzo K\"uchler, Tjonnie Li, Alex Lupsasca, Emil Martinec, Hector Olivares, Paolo Pani, Thomas Van Riet, Bart Ripperda, Seppe Staelens, and Nick Warner for discussions.
B.V. would like to thank Tom Lemmens, Marina Martinez-Rodriguez for work on philosophically related projects. B.V. would like to thank his many colleagues and friends he had the pleasure to work with over the past years.

The work of B.V. and D.R.M. is supported by Odysseus grant G0H9318N of FWO Vlaanderen.
This work is also partially supported by the KU Leuven C1 grant ZKD1118 C16/16/005.

\phantomsection
\bibliography{fb_manifesto}

\end{document}